\documentclass[aps,prc,superscriptaddress,floatfix,nofootinbib,preprint]{revtex4}
\usepackage{color}
\usepackage{dcolumn}
\usepackage{graphicx}

\newcommand{\s}{\sigma}
\newcommand{\pp}{\pi^+\pi^-}

\begin{document}

\title{Search for the QCD critical point in nuclear collisions at the CERN SPS}

%============================================================================
\affiliation{NIKHEF, Amsterdam, Netherlands}
\affiliation{Department of Physics, University of Athens, Athens, Greece}
\affiliation{Comenius University, Bratislava, Slovakia}
\affiliation{KFKI Research Institute for Particle and Nuclear Physics,
             Budapest, Hungary}
\affiliation{MIT, Cambridge, Massachusetts, USA}
\affiliation{Henryk Niewodniczanski Institute of Nuclear Physics,
             Polish Academy of Science, Cracow, Poland}
\affiliation{Gesellschaft f\"{u}r Schwerionenforschung (GSI),
             Darmstadt, Germany}
\affiliation{Joint Institute for Nuclear Research, Dubna, Russia}
\affiliation{Fachbereich Physik der Universit\"{a}t, Frankfurt, Germany}
\affiliation{CERN, Geneva, Switzerland}
\affiliation{Institute of Physics, Jan Kochanowski University, Kielce, Poland}
\affiliation{Fachbereich Physik der Universit\"{a}t, Marburg, Germany}
\affiliation{Max-Planck-Institut f\"{u}r Physik, Munich, Germany}
\affiliation{Institute of Particle and Nuclear Physics, Charles
             University, Prague, Czech Republic}
\affiliation{Nuclear Physics Laboratory, University of Washington,
             Seattle, Washington, USA}
\affiliation{Atomic Physics Department, Sofia University St.~Kliment
             Ohridski, Sofia, Bulgaria}
\affiliation{Institute for Nuclear Research and Nuclear Energy, Sofia, Bulgaria}
\affiliation{Department of Chemistry, Stony Brook University (SUNYSB), Stony Brook, New York, USA}
\affiliation{Institute for Nuclear Studies, Warsaw, Poland}
\affiliation{Institute for Experimental Physics, University of Warsaw,
             Warsaw, Poland}
\affiliation{Faculty of Physics, Warsaw University of Technology, Warsaw, Poland}
\affiliation{Rudjer Boskovic Institute, Zagreb, Croatia}

%============================================================================
\author{T.~Anticic}
\affiliation{Rudjer Boskovic Institute, Zagreb, Croatia}
\author{B.~Baatar}
\affiliation{Joint Institute for Nuclear Research, Dubna, Russia}
\author{D.~Barna}
\affiliation{KFKI Research Institute for Particle and Nuclear Physics,
             Budapest, Hungary}
\author{J.~Bartke}
\affiliation{Henryk Niewodniczanski Institute of Nuclear Physics,
             Polish Academy of Science, Cracow, Poland}
\author{L.~Betev}
\affiliation{CERN, Geneva, Switzerland}
\author{H.~Bia{\l}\-kowska}
\affiliation{Institute for Nuclear Studies, Warsaw, Poland}
\author{C.~Blume}
\affiliation{Fachbereich Physik der Universit\"{a}t, Frankfurt, Germany}
\author{B.~Boimska}
\affiliation{Institute for Nuclear Studies, Warsaw, Poland}
\author{M.~Botje}
\affiliation{NIKHEF, Amsterdam, Netherlands}
\author{J.~Bracinik}
\affiliation{Comenius University, Bratislava, Slovakia}
\author{P.~Bun\v{c}i\'{c}}
\affiliation{CERN, Geneva, Switzerland}
\author{V.~Cerny}
\affiliation{Comenius University, Bratislava, Slovakia}
\author{P.~Christakoglou}
\affiliation{NIKHEF, Amsterdam, Netherlands}
\author{P.~Chung}
\affiliation{Department of Chemistry, Stony Brook University (SUNYSB),
             Stony Brook, New York, USA}
\author{O.~Chvala}
\affiliation{Institute of Particle and Nuclear Physics, Charles
             University, Prague, Czech Republic}
\author{J.~G.~Cramer}
\affiliation{Nuclear Physics Laboratory, University of Washington,
             Seattle, Washington, USA}
\author{P.~Dinkelaker}
\affiliation{Fachbereich Physik der Universit\"{a}t, Frankfurt, Germany}
\author{V.~Eckardt}
\affiliation{Max-Planck-Institut f\"{u}r Physik, Munich, Germany}
\author{Z.~Fodor}
\affiliation{KFKI Research Institute for Particle and Nuclear Physics,
             Budapest, Hungary}
\author{P.~Foka}
\affiliation{Gesellschaft f\"{u}r Schwerionenforschung (GSI),
             Darmstadt, Germany}
\author{V.~Friese}
\affiliation{Gesellschaft f\"{u}r Schwerionenforschung (GSI),
             Darmstadt, Germany}
\author{M.~Ga\'zdzicki}
\affiliation{Fachbereich Physik der Universit\"{a}t, Frankfurt, Germany}
\affiliation{Institute of Physics, Jan Kochanowski University, Kielce, Poland}
\author{V.~Genchev}
\affiliation{Institute for Nuclear Research and Nuclear Energy, Sofia,
             Bulgaria}
\author{K.~Grebieszkow}
\affiliation{Faculty of Physics, Warsaw University of Technology,
             Warsaw, Poland}
\author{C.~H\"{o}hne}
\affiliation{Gesellschaft f\"{u}r Schwerionenforschung (GSI),
             Darmstadt, Germany}
\author{K.~Kadija}
\affiliation{Rudjer Boskovic Institute, Zagreb, Croatia}
\author{A.~Karev}
\affiliation{Max-Planck-Institut f\"{u}r Physik, Munich, Germany}
\author{V.~I.~Kolesnikov}
\affiliation{Joint Institute for Nuclear Research, Dubna, Russia}
\author{M.~Kowalski}
\affiliation{Henryk Niewodniczanski Institute of Nuclear Physics,
             Polish Academy of Science, Cracow, Poland}
\author{M.~Kreps}
\affiliation{Comenius University, Bratislava, Slovakia}
\author{A.~Laszlo}
\affiliation{KFKI Research Institute for Particle and Nuclear Physics,
             Budapest, Hungary}
\author{R.~Lacey}
\affiliation{Department of Chemistry, Stony Brook University (SUNYSB),
             Stony Brook, New York, USA}
\author{M.~van~Leeuwen}
\affiliation{NIKHEF, Amsterdam, Netherlands}
\author{B.~Lungwitz}
\affiliation{Fachbereich Physik der Universit\"{a}t, Frankfurt, Germany}
\author{M.~Makariev}
\affiliation{Institute for Nuclear Research and Nuclear Energy, Sofia, Bulgaria}
\author{A.~I.~Malakhov}
\affiliation{Joint Institute for Nuclear Research, Dubna, Russia}
\author{M.~Mateev}
\affiliation{Atomic Physics Department, Sofia University St.~Kliment
             Ohridski, Sofia, Bulgaria}
\author{G.~L.~Melkumov}
\affiliation{Joint Institute for Nuclear Research, Dubna, Russia}
\author{M.~Mitrovski}
\affiliation{Fachbereich Physik der Universit\"{a}t, Frankfurt, Germany}
\author{St.~Mr\'owczy\'nski}
\affiliation{Institute of Physics, Jan Kochanowski University, Kielce, Poland}
\author{V.~Nicolic}
\affiliation{Rudjer Boskovic Institute, Zagreb, Croatia}
\author{G.~P\'{a}lla}
\affiliation{KFKI Research Institute for Particle and Nuclear Physics,
             Budapest, Hungary}
\author{A.~D.~Panagiotou}
\affiliation{Department of Physics, University of Athens, Athens, Greece}
\author{A.~Petridis}
\thanks{Deceased.}
\affiliation{Department of Physics, University of Athens, Athens, Greece}
\author{W.~Peryt}
\affiliation{Faculty of Physics, Warsaw University of Technology,
             Warsaw, Poland}
\author{M.~Pikna}
\affiliation{Comenius University, Bratislava, Slovakia}
\author{J.~Pluta}
\affiliation{Faculty of Physics, Warsaw University of Technology,
             Warsaw, Poland}
\author{D.~Prindle}
\affiliation{Nuclear Physics Laboratory, University of Washington,
             Seattle, Washington, USA}
\author{F.~P\"{u}hlhofer}
\affiliation{Fachbereich Physik der Universit\"{a}t, Marburg, Germany}
\author{R.~Renfordt}
\affiliation{Fachbereich Physik der Universit\"{a}t, Frankfurt, Germany}
\author{C.~Roland}
\affiliation{MIT, Cambridge, Massachusetts, USA}
\author{G.~Roland}
\affiliation{MIT, Cambridge, Massachusetts, USA}
\author{M.~Rybczy\'nski}
\affiliation{Institute of Physics, Jan Kochanowski University, Kielce, Poland}
\author{A.~Rybicki}
\affiliation{Henryk Niewodniczanski Institute of Nuclear Physics,
             Polish Academy of Science, Cracow, Poland}
\author{A.~Sandoval}
\affiliation{Gesellschaft f\"{u}r Schwerionenforschung (GSI),
             Darmstadt, Germany}
\author{N.~Schmitz}
\affiliation{Max-Planck-Institut f\"{u}r Physik, Munich, Germany}
\author{T.~Schuster}
\affiliation{Fachbereich Physik der Universit\"{a}t, Frankfurt, Germany}
\author{P.~Seyboth}
\affiliation{Max-Planck-Institut f\"{u}r Physik, Munich, Germany}
\author{F.~Sikl\'{e}r}
\affiliation{KFKI Research Institute for Particle and Nuclear Physics,
             Budapest, Hungary}
\author{B.~Sitar}
\affiliation{Comenius University, Bratislava, Slovakia}
\author{E.~Skrzypczak}
\affiliation{Institute for Experimental Physics, University of Warsaw,
             Warsaw, Poland}
\author{M.~Slodkowski}
\affiliation{Faculty of Physics, Warsaw University of Technology,
             Warsaw, Poland}
\author{G.~Stefanek}
\affiliation{Institute of Physics, Jan Kochanowski University, Kielce, Poland}
\author{R.~Stock}
\affiliation{Fachbereich Physik der Universit\"{a}t, Frankfurt, Germany}
\author{H.~Str\"{o}bele}
\affiliation{Fachbereich Physik der Universit\"{a}t, Frankfurt, Germany}
\author{T.~Susa}
\affiliation{Rudjer Boskovic Institute, Zagreb, Croatia}
\author{M.~Szuba}
\affiliation{Faculty of Physics, Warsaw University of Technology,
             Warsaw, Poland}
\author{D.~Varga}
\affiliation{KFKI Research Institute for Particle and Nuclear Physics,
             Budapest, Hungary}
\author{M.~Vassiliou}
\affiliation{Department of Physics, University of Athens, Athens, Greece}
\author{G.~I.~Veres}
\affiliation{KFKI Research Institute for Particle and Nuclear Physics,
             Budapest, Hungary}
\author{G.~Vesztergombi}
\affiliation{KFKI Research Institute for Particle and Nuclear Physics,
             Budapest, Hungary}
\author{D.~Vrani\'{c}}
\affiliation{Gesellschaft f\"{u}r Schwerionenforschung (GSI),
             Darmstadt, Germany}
\author{Z.~W{\l}odarczyk}
\affiliation{Institute of Physics, Jan Kochanowski University, Kielce, Poland}

%============================================================================

\collaboration{NA49 Collaboration}
\noaffiliation
\author{N.~G.~Antoniou}
\affiliation{Department of Physics, University of Athens, Athens, Greece}
\author{F.~K.~Diakonos}
\affiliation{Department of Physics, University of Athens, Athens, Greece}
\author{G.~Mavromanolakis}
\affiliation{Cavendish Laboratory, University of Cambridge, Cambridge CB3 0HE, UK}

%============================================================================

\begin{abstract}

Pion production in nuclear collisions at the SPS is investigated with the aim to
search, in a restricted domain of the phase diagram, for power-laws in the
behavior of correlations which are compatible with critical QCD. We have
analyzed interactions of nuclei of different size (p+p, C+C, Si+Si, Pb+Pb) at
158$A$~GeV adopting, as appropriate observables, scaled factorial moments in a
search for intermittent fluctuations in transverse dimensions. The analysis is performed for $\pi^+\pi^-$ pairs with
invariant mass very close to the two-pion threshold.
In this sector one may capture critical fluctuations of the sigma component in a hadronic medium,
even if the $\sigma$-meson has no well defined vacuum state. It turns out that for the Pb+Pb system the proposed analysis technique cannot be applied without entering the invariant mass region with strong Coulomb correlations. As a result the treatment becomes inconclusive in this case. Our results for the other systems indicate the presence of power-law fluctuations in the freeze-out state of Si+Si approaching in size the prediction of critical QCD.

\end{abstract}
\maketitle

\date{\today}

\section[roman]{Introduction}

\hspace*{0.3cm} The experiments with nuclei at the CERN SPS are dedicated to the search for evidence of a deconfined state of quarks and gluons at high temperatures,
separated from conventional hadronic matter by a critical line in the two-dimensional phase diagram $(\mu_B,T)$. Close to this line significant fluctuations associated with the quark-hadron phase transition occur. In principle the experimental study of these fluctuations becomes feasible in a class of nuclear collisions whose reaction volume freezes out in this area. The underlying theory of strongly interacting matter (QCD) suggests that, across the critical line, the phase transition is of first order, at least for large values of $\mu_B$, whereas at zero chemical potential $(\mu_B=0)$ the transition becomes a smooth crossover \cite{Wilcz00}. This picture implies that there is a point in the phase diagram, located at nonzero baryonic density, where the first-order transition line stops (endpoint). On the basis of general considerations this endpoint is characterized by a second-order phase transition and it becomes a distinct property of strongly interacting matter (QCD critical endpoint). The related critical phenomena give rise to density fluctuations which obey appropriate power-laws, specified by the critical exponents of this transition
\cite{Wilcz00,Ant05}. This critical endpoint of QCD matter is the remnant of the chiral phase transition, and its existence becomes the fact that light quarks acquire a small but nonzero mass which breaks chiral symmetry of strong interactions explicitly \cite{Wilcz00}.\\
\vspace*{0.1cm}

\hspace*{0.3cm} Quantitatively, the power-laws of QCD matter at criticality describe the density fluctuations of zero mass $\s$-particles produced abundantly in a nuclear collision at the critical point \cite{Wilcz00,Ant05}. In particular, critical fluctuations of the $\s$-field ($(\delta \s)^2 \approx \langle \sigma^2 \rangle$) in the transverse configuration plane, with respect to the beam axis, are characterized by a power-law at large distances of the form \cite{Ant05}: $(\delta \s)^2 \sim \vert \vec{x}_{\perp} \vert^{d_F-2}$ where $d_F=2 (\delta -1)/(\delta +1)$ and $\delta$ is the isothermal critical exponent of hot QCD matter $(\delta \approx 5)$. Any experimental attempt to verify this power-law as a signature of the QCD critical point must rely on the corresponding effect in momentum space in analogy to conventional matter at criticality, where a similar power-law in configuration space leads to the spectacular observable effect of critical opalescence \cite{Stan87} in momentum space (scattering of long wavelength light). In fact, the correlator in momentum space $\langle n_{\vec{p}}~ n_{\vec{p}+\vec{k}} \rangle$ associated with the occupation number of sigmas in transverse momentum states, obeys in the case of critical QCD matter a power-law for small $\vert \vec{k} \vert$ of the form $\langle n_{\vec{p}}~ n_{\vec{p}+\vec{k}} \rangle \sim \vert \vec{k} \vert^{-d_F}$ which reflects the critical nature of density fluctuations (of sigmas) in configuration space. The singularity in the limit $\vert \vec{k} \vert \to 0$ of the correlator and the associated intermittency pattern provide the basic observables in the search for measurable effects related to the critical behavior of QCD.\\
\vspace*{0.1cm}

\hspace*{0.3cm} The sigma states in this approach are identified with $\pi^+\pi^-$ pairs of invariant mass distributed near the two-pion threshold according to a spectral enhancement of the form $\rho_{\sigma}(m_{\pi^+\pi^-}) \sim \left( 1 - \frac{4 m_{\pi}^2}{m_{\pi^+\pi^-}^2} \right)^{-1/2}$
\cite{Kunihiro99}. In this restricted domain of the phase space the $\pi^+\pi^-$ system has the quantum numbers of the sigma field ($I=J=0$), whereas the singularity of $\rho_{\sigma}(m_{\pi^+\pi^-})$ at threshold is related to partial restoration of chiral symmetry \cite{Kunihiro99} as we approach the critical point in a hadronic medium (finite temperature and baryon density). With this prescription one expects, by studying the behavior of $\pi^+\pi^-$ pairs (near threshold) in the freeze-out states of nuclear collisions, to be able to capture the properties of the sigma field (order parameter) at the critical point. It is of interest to note that the spectral function of sigma in a thermal environment and near the two-pion threshold is based on general principles (partial restoration of chiral symmetry) and is not affected by the controversial issue of $\sigma$-meson (a broad resonance) in vacuum $(T=0, \rho_B=0)$ which remains an open question in hadronic physics \cite{Kunihiro99,CHAOS}.\\
\vspace*{0.1cm}

\hspace*{0.3cm} In this work an experimental search for the QCD critical point is performed along these lines in the freeze-out environment of nuclear collisions at the CERN SPS. The investigation is based on the NA49 measurements of multipion production in central collisions at
158$A$~GeV in a series of systems of different sizes (p+p, C+C, Si+Si, Pb+Pb). The motivation for this search comes from theoretical studies suggesting that the QCD critical point is likely to be within reach at the SPS energies \cite{Fodor04}. Moreover, the anomalies in the energy dependence of hadron production reported recently \cite{NA49041} indicate that a first-order quark-hadron phase transition starts in central Pb+Pb collisions at beam energies around 30$A$~GeV \cite{NA49042}. As a consequence, the second-order critical endpoint is likely to be located in a region of lower baryon chemical potential which, presumably, can be reached by varying the system size with energies close to the highest SPS energy (158$A$~GeV).\\
\vspace*{0.1cm}

\hspace*{0.3cm} The observables in this search are chosen to be sensitive to the power-laws of the correlation functions which are valid at the critical point of QCD matter. To this end we follow the proposal in ref.~\cite{Ant05} where two-dimensional scaled factorial moments of order $p$, $F_p(M)$, defined in small transverse momentum 2D cells $\delta S$ ($\delta S \sim M^{-2}$ where $M^2$ is the number of cells),
are suggested as the most suitable observables in the sigma mode ($\pp$ pairs) near the two-pion threshold. In accordance with our previous discussion, the counterpart in momentum space of the QCD critical power-law in configuration space is the phenomenon of intermittency \cite{Bial86} with a linear spectrum of indices (critical intermittency): $F_p(M) \sim M^{2 \phi_2 (p-1)}$ ($M \gg 1$), $\phi_2 = (\delta -1)/(\delta +1)$, directly observable in the reconstructed sector of sigmas \cite{Ant05}.\\
\vspace*{0.1cm}

\hspace*{0.3cm} In section II we describe the experiment and the pion data sets obtained by imposing appropriate event and track cuts. In section III the method of analysis is explained with emphasis on the reconstruction of the $\s$-sector near the two-pion threshold. In particular it is shown that the elimination of the combinatorial background from the correlation of pion pairs ($\pp$) can be achieved to a large extent by a suitable subtraction of the factorial moments of mixed events \cite{Ant05}. In addition Coulomb correlated $\pi^+\pi^-$ pairs are excluded by imposing appropriate kinematical cuts in the invariant mass. In section IV we apply this method to data sets from p+p, C+C, Si+Si, Pb+Pb collisions at 158$A$~GeV and perform a systematic search for a power-law behavior of factorial moments in the sigma mode. Moreover, a study is made of the compatibility of the results with the predicted behavior of critical QCD matter. A comparison is also performed with conventional Monte Carlo (HIJING) and critical Monte Carlo (CMC) predictions. Finally, in section V our findings are summarized and discussed together with the limitations of the method and the prospects for further investigations in current and future experiments concerning the QCD critical point.\\

\section[roman]{Experimental setup and pion production data at the SPS}

\hspace*{0.3cm} The NA49 experimental setup \cite{Afan99} is shown in Fig.~1. %(\ref{exsetup}).
The main detectors of the experiment are four large-volume time projection chambers (TPCs). Two of these, the vertex TPCs (VTPC-1 and VTPC-2), are located in the magnetic field of two superconducting dipole magnets. This allows separation of positively and negatively charged tracks and a measurement of the particle momenta. The other two TPCs (MTPC-L and MTPC-R), positioned downstream of the magnets, are optimized for precise measurement of the ionization energy loss $dE/dx$ which is used for the determination of the particle masses. Additional information on the particle masses is provided by two time-of-flight (TOF) detector arrays which are placed behind the MTPCs. The centrality of the collisions is determined by a calorimeter (VCAL) which measures the energy of the projectile spectators. To cover only the spectator region the geometrical acceptance of the VCAL was adjusted by a proper setting of a collimator (COLL) \cite{Afan99,Appel98}. The beam position detectors (BPD-1, BPD-2, and BPD-3) are used to determine the $x$ and $y$ coordinates of each beam particle at the target. Alternatively, the main vertex position is reconstructed as the common intersection point of reconstructed tracks. A detailed description of the NA49 setup and tracking software can be found in ref.~\cite{Afan99}.\\

\begin{figure}[htbp]
\centerline{\includegraphics[width=17 cm,height=5.5 cm]{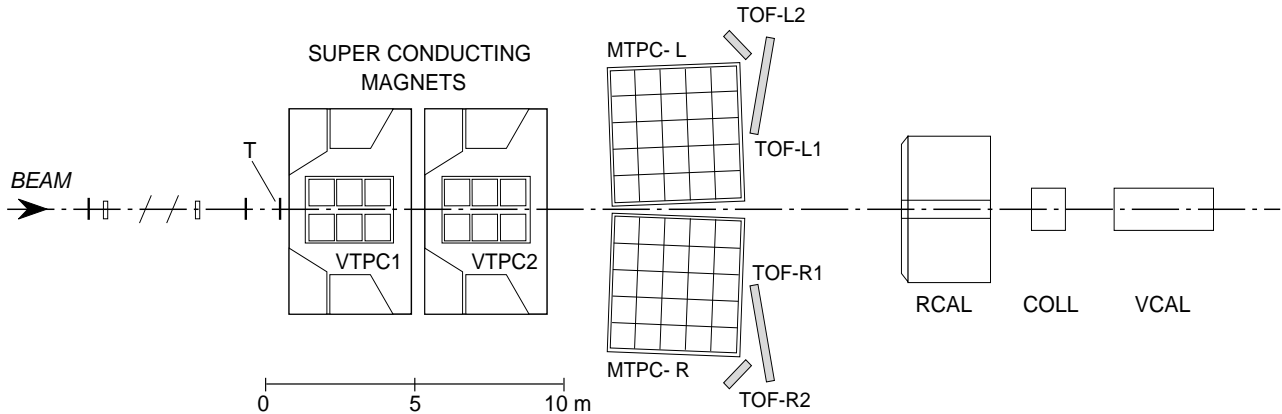}}
\caption{The NA49 experimental setup.}
\label{fig:fig1}
\end{figure}
%\begin{figure}[h]
%\begin{center}
%\mbox{\epsfig{figure=exsetup.eps,width=11cm,angle=0}}
%\caption{Figure 1}
%\label{exsetup}
%\end{center}
%\end{figure}

The targets are C ($561$~mg/cm$^2$), Si ($1170$~mg/cm$^2$) disks and a Pb ($224$~mg/cm$^2$) foil for ion collisions and a liquid hydrogen cylinder (length $20$~cm) for proton interactions. They are positioned about $80$~cm upstream from VTPC-1. A total of 33689 C+C, 17053 Si+Si, 30000 Pb+Pb and 408708 p+p events after all necessary rejections with respect to beam charge and vertex position were selected from 1998, 1998, 1996 and 1998 run periods, respectively. For the Si+Si and C+C systems all of the 10\% most central events were used. The Pb+Pb events analyzed here were selected as the 5 \% most central ones. Further selection cuts were applied at the track level. In order to reject double tracks or partially reconstructed tracks a requirement is set on the number of reconstructed points in the TPCs between 20 and 235 and a ratio of reconstructed over potential points above 0.5.
Tracks were selected in a momentum band of $3 - 50$~GeV/c in order to allow pion identification by using $dE/dx$ information recorded in the TPC.
A cut of 1 sigma around the momentum dependent pion peak of the $dE/dx$ distribution was applied, reducing the residual background of other particle types to a very low level (less than $0.3$\%). This contamination probability is estimated through the overlap of the tails of the kaon and baryon $dE/dx$ distributions with the 1 sigma region around the peak of the pion $dE/dx$ distribution.\\

\section[roman]{Optimal reconstruction of the $\pi^+\pi^-$ critical sector using CMC simulations}

\hspace*{0.3cm} The guideline for the development of an efficient algorithm for the reconstruction  of the sigma sector and its density fluctuations
is obtained through the analysis of events generated by the Critical Monte-Carlo (CMC) code \cite{Ant05}.
The sigma sector of the CMC events is characterized by self-similar density fluctuations corresponding to a fractal set with dimension $D_f=2/3$ in transverse momentum space. These fluctuations are not transferred to the daughter pions directly due to decay kinematics. Using the fact that the critical sigma sector as well as its geometrical properties are known in the CMC events one has a measure for the efficiency of a reconstruction algorithm using the observed momenta of pions of opposite charge $(\pi^+,\pi^-)$. In such a scheme the direct observation of the critical sigmas is not possible as they are hidden in a large background of $\pi^+\pi^-$ pairs, formed from pions originating from two different sigmas. However even if the sigma itself is not observable its self-similar fluctuations in the isoscalar sector of pions can be revealed by using a suitable algorithm which is developed and extensively described in \cite{Ant05}. Its basic steps are summarized in the following:
\begin{enumerate}
\item{For each event in a given data set, consisting of identified positive and negative pions, all possible pairs $\pi^+\pi^-$ with invariant mass in a small kinematical window of size $\Delta \epsilon=\epsilon_2-\epsilon_1$ above the two-pion threshold are formed:
\begin{equation}
(2 m_{\pi} +\epsilon_1)^2 \leq (p_{\pi^+} + p_{\pi^-})^2 \leq (2 m_{\pi} + \epsilon_2)^2
\label{eq:epswin}
\end{equation}
with $\epsilon_i \ll 2 m_{\pi}, i=1,2$. The sum of the momenta of $\pi^+$ and $\pi^-$, constituting a pair, determines the momentum of the corresponding dipion: $\vec{p}_{\pi\pi}=\vec{p}_{\pi^+} + \vec{p}_{\pi^-}$. The parameters $\epsilon_1$, $\epsilon_2$ in
eq.~(\ref{eq:epswin}) can be used to displace the kinematical window of analysis along the dipion invariant-mass axis as well as to modify its size. Such a displacement can be useful in order to avoid the presence of Coulomb correlations in the considered data set. In fact if $\epsilon_1$ fulfils the constraint $\epsilon_1 \stackrel{>}{\sim}5$~MeV ($Q_{inv} \stackrel{>}{\sim} 53$~MeV) the domain of Coulomb correlated $\pi^+\pi^-$ pairs is excluded from the considered kinematical window in the reconstructed dipion sector \cite{Sey98}. After this filtering procedure, only $\pi^+\pi^-$ pairs satisfying the condition
(\ref{eq:epswin}) are retained for further analysis.}
\item{The pion pairs produced from the previous step are used to analyze the density fluctuations in the isoscalar sector, employing as a suitable tool the $2D$ transverse momentum factorial moments \cite{Bial86}:
\begin{equation}
F_p(M)=\frac{\langle \displaystyle{\frac{1}{M^2}\sum_{i=1}^{M^2}} n_i(n_i-1)..(n_i-p+1) \rangle}
{\langle \displaystyle{\frac{1}{M^2}\sum_{i=1}^{M^2}} n_i \rangle^p}
\label{eq:facmom}
\end{equation}
with $M^2$ the number of cells in transverse momentum space and $n_i$ the number of reconstructed dipions in the $i$-th cell. A power-law dependence $F_2 \sim (M^2)^{s_2}$ for large $M$ ($s_2$ is the corresponding intermittency exponent) indicates the presence of self-similar fluctuations in transverse momenta. However before judging the power-law behavior of the factorial moments one has to eliminate the effect of non-critical dipions present in the events (step 1) as a result of combinatorial background (see step 3). }
\item{The second factorial moment calculated in step 2 is built by the density-density correlation between any two dipions $(\pi^+\pi^-)$ consisting both of critical sigmas (critical dipions) and non-critical dipions. The dominant background in $F_2(M)$ consists of density-density correlations between two non-critical dipions and must be subtracted in order to reveal the self-similar correlations between two critical dipions, if they exist.
Practically, to achieve this subtraction we assume that the number of dipions $n_i$ in the $i$-th transverse momentum space cell occurring in
(\ref{eq:facmom}) is decomposed in signal $n_{\sigma,i}$ (critical sigmas) and background $n_{b,i}$ (non-critical dipions) which are (approximately) statistically independent:
\begin{eqnarray}
n_i&=&n_{\sigma,i}+n_{b,i} \nonumber \\
\langle n_i (n_i-1) \rangle &=& \langle n_{\sigma,i} (n_{\sigma,i}-1) \rangle + \langle n_{b,i}(n_{b,i}-1) \rangle + 2 \langle n_{\sigma,i} n_{b,i} \rangle
\nonumber \\
\langle n_{\sigma,i} n_{b,i} \rangle &\approx& \langle n_{\sigma,i} \rangle \langle n_{b,i} \rangle
\label{eq:decomp}
\end{eqnarray}
Inserting (\ref{eq:decomp}) in (\ref{eq:facmom}) and assuming further that the background is efficiently simulated by mixed events we obtain the following formula for the subtracted factorial moment $\Delta F_2$:
\begin{equation}
\Delta F_2(M)=F_2(M)-x_M^2 F_2^{(m)}(M) -2 x_M (1-x_M)~~~~~;~~~~x_M=\frac{\langle n^{(m)} \rangle_M}{\langle n \rangle_M}
\label{eq:correlator}
\end{equation}
where $F_2^{(m)}(M)$ is the second factorial moment in transverse momentum space calculated using dipions originating from mixed events while
$\langle n \rangle_M$ and $\langle n^{(m)} \rangle_M$ are the mean numbers of reconstructed dipions in a cell obtained from data and mixed events, respectively. In $\Delta F_2$ a large part of the non-critical dipions are expected to be eliminated and the fluctuations carried by the critical $\pi^+\pi^-$ pairs can be revealed to a large extent. Therefore in the following analysis we will exclusively use the correlator
(\ref{eq:correlator}) in our search for critical fluctuations. For the simulated CMC events or for an $A+A$ system freezing out exactly at the critical point, $\Delta F_2$ is expected to possess a power-law behavior: $\Delta F_2 \sim (M^2)^{\phi_2}$ with the critical index $\phi_2=2/3$ determined by the universality class of the transition \cite{Ant05}.  It must be noted that the effect we are looking for is associated with small momentum scales. Therefore we perform the intermittency analysis using for the number of cells the condition $M^2 \geq 2000$. In this way we avoid in real data the influence of possible structures at large momentum  scales on the determination of the critical index $\phi_2$. In practice $\phi_2$ is obtained through a power-law fit of $\Delta F_2$ while the quality parameters $\chi^2$ and $R^2$ are used to verify its validity\footnote{The coefficient of determination $R^2$ is defined through \cite{Eve02}:
$$R^2=\frac{(<y \tilde{y}>-<y> <\tilde{y}>)^2}{(<y^2>-<y>^2)(<\tilde{y}^2>-<\tilde{y}>^2)}$$
where $y$ are the experimentally observed values for a given observable $Y$ while $\tilde{y}$ are the corresponding values of the fitting function.}. The resulting values $\chi^2 /dof < 1$ (dof = number of degrees of freedom) and $R^2$ approaching $1$ indicate that the dependence of $\Delta F_2$ on $M^2$ is consistent with a power-law behavior. The proposed subtraction method is an approximate one, its validity is discussed below based on the results
of the Critical Monte Carlo.}
\item{It is known in the literature that the intermittency exponents are sensitive to the multiplicity of the analyzed events \cite{deWolf96}. As a consequence, in order to compare the results of intermittency analysis in systems having different sizes $A$, and therefore different charged pion
multiplicity, we have to remove this bias. This is achieved if the mean multiplicity of reconstructed dipions is as closely as possible the same for the various systems. To this end one should tune appropriately the size of the kinematical window in (\ref{eq:epswin}). One can show \cite{Ant05} that when the mean number of  reconstructed dipions decreases the relative weight of critical to non-critical dipions in the reconstructed events increases. However, decreasing the multiplicity in general worsens the statistics and an optimization with respect to the choice of $\Delta \epsilon = \epsilon_2 - \epsilon_1$ in (\ref{eq:epswin}) is in order depending on the particular data set considered. This optimization procedure has also to take into account the limited experimental invariant-mass resolution $\delta m$ which restricts the kinematical window $\Delta \epsilon$ accordingly
($\Delta \epsilon > \delta m$).}
\item{The critical sigmas, as explained in the introduction, are expected to have an invariant-mass distribution peaked at the two-pion threshold. Since the number of critical sigmas within the kinematical domain (\ref{eq:epswin}) is crucial for the efficient reconstruction of the power-law discussed above, one has to look for an interval with enhanced content in critical sigmas, scanning for this purpose an extended kinematical range of dipion invariant mass. This is achieved by varying $\epsilon_1$ in eq.~(\ref{eq:epswin}). In order to suppress the influence of other hadronic resonances one should restrict the position of the kinematical window to a narrow region just above the two-pion threshold. A safe choice would be, for example, $\epsilon_2 \leq 70$~MeV \cite{Rapp03}.}
\end{enumerate}
\vspace*{0.1cm}

\hspace*{0.3cm} The efficiency of the above algorithm can be tested by applying it to data sets consisting of CMC critical events. In this case, the observable pionic sector is produced through the decay of the critical sigmas, generated by the CMC code, into pions. At this point we need to incorporate in the Critical Monte Carlo, along with the universal power-laws, the fact that the $\sigma$-activity just above the two-pion threshold is associated with the partial restoration of chiral symmetry as the system approaches the critical point \cite{Kunihiro99}. In particular the
$\pi^+\pi^-$ invariant-mass distribution follows a characteristic spectral enhancement in the $\sigma$-mode \cite{Kunihiro99} at the $2 m_{\pi}$ threshold, $\rho_{\sigma}(m_{\pi^+\pi^-}) \sim \left( 1 - \frac{4 m_{\pi}^2}{m_{\pi^+\pi^-}^2} \right)^{-1/2}$ which plays a crucial role in the reconstruction of self consistent critical $\sigma$-correlations in our treatment.\\

\begin{figure}[htbp]
\centerline{\includegraphics[width=18 cm,height=14 cm]{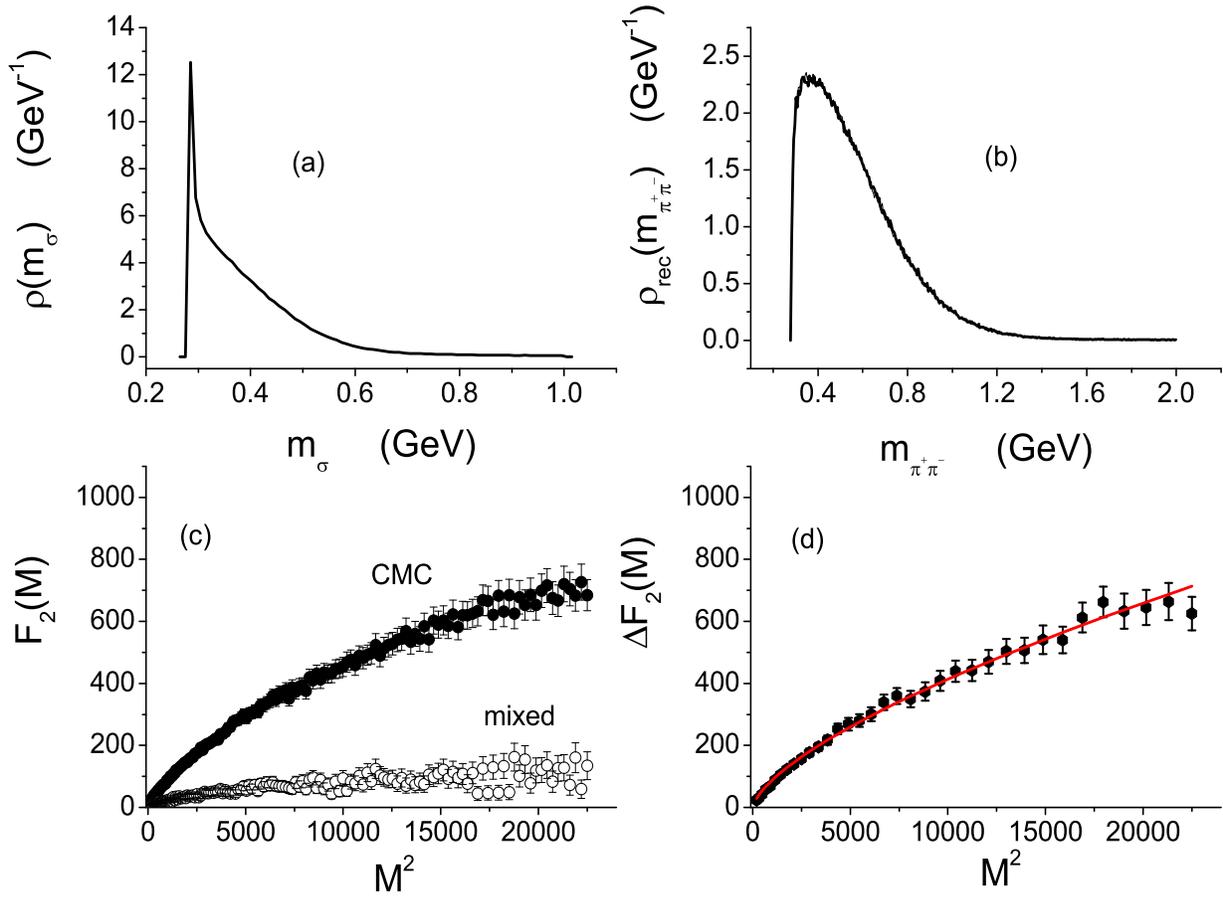}}
\caption{(a) The spectral density of sigmas at the critical point according to ref.~\cite{Kunihiro99} and (b) the corresponding dipion invariant-mass distribution resulting from reconstruction using simulated CMC events. In (c) we show the second factorial moments $F_2$ (full circles) and $F_2^{(m)}$ (open circles) in transverse momentum space of the CMC dipion sector using dipions with invariant mass in the kinematical window $[280,280.6]$~MeV (optimal reconstruction of critical fluctuations) and in (d) we show the corresponding subtracted moment $\Delta F_2$. The solid line is the result of the power-law fit leading to $\phi_2=0.67 \pm 0.01$.}
\label{fig:fig2abcd}
\end{figure}

\vspace*{0.1cm}
\hspace*{0.3cm} In Fig.~2(a-d) we illustrate how the above spectrum leads to the reconstruction of the critical index $\phi_2$ in the immediate neighborhood of the $2 m_{\pi}$ threshold. It is seen that although the spike in the invariant-mass spectrum of the sima (Fig.~2a) is not observable in the reconstructed dipion sector (Fig.~2b), owing to combinatorics and finite statistics, the fluctuations of the underlying critical sigmas can be revealed by intermittency analysis in transverse momentum space, if the invariant-mass window of the reconstructed $\pi^+\pi^-$ pairs is chosen to be located close to the $2 m_{\pi}$ threshold (Fig.~2c,d). In Fig.~2c we present the second factorial moments $F_2$ (full circles) and $F_2^{(m)}$ (open circles) in transverse momentum space of reconstructed dipions in the case of optimal reconstruction of critical fluctuations using in (1) the kinematical window $[280,280.6]$~MeV. The mean multiplicity of reconstructed dipions in this interval is 1.6. The subtracted moment $\Delta F_2$ is shown
in Fig.~2d. The solid line is the result of the power-law fit. The reconstruction of the critical fluctuations measured through $\phi_2$ leads to the theoretically predicted value ($\phi_2=0.67 \pm 0.01$) with $R^2 = 1$ ($\chi^2 \approx 0.4$).\\
\vspace*{0.1cm}

\hspace*{0.3cm} We have also explored how the results of the analysis change as we get off the optimal scenario by changing the location or the size of the dipion invariant-mass window used in the reconstruction as well as by contaminating the ensemble of pions originating from the decay of critical sigmas with pions produced from a random source. The results of this analysis are shown in Fig.~3(a-d). More specifically in Fig.~3a we show the decrease of the $\phi_2$ value as well as the reduction of the quality of the corresponding power-law fit (increasing error bars), when the location of the kinematical window used in the analysis is placed at increasing distance from the two-pion threshold. Each window used in Fig.~3a has an appropriate size, varying in the range $[0.6,1]$~MeV, so that the corresponding mean number of pions is $\langle n_{\pi^+\pi^-} \rangle_{\Delta \epsilon} \approx 2$. For the decay of the sigmas into pions the spectral density shown in Fig.~2a is used. The vertical solid line at $285$~MeV is drawn to indicate the dipion invariant mass value below which the Coulomb correlations are strong. In Fig.~3b we show the correlator $\Delta F_2(M)$ for the dipion invariant mass window $[285,286]$~MeV in double logarithmic scale. The dashed line indicates the associated linear fit used for the estimation of $\phi_2$. To determine the dependence of the results of the fluctuation analysis on the size of the kinematical window we have calculated $\phi_2$ employing in the reconstruction dipion invariant-mass windows $[280$~MeV,$280$~MeV $+ \Delta \epsilon]$ with increasing size $\Delta \epsilon$. The exact profile of this dependence is expected to be sensitive to the mean multiplicity of initial critical sigmas which decay into pions forming the ensemble of CMC events. When the mean number of initial sigmas gets smaller, the corresponding combinatorial background (non-critical dipions) decreases as well. Since the number of non-critical dipions $\langle n_b \rangle$ depends quadratically on the mean number of decaying sigmas $\langle n_{\sigma} \rangle$ the relative weight $\frac{\langle n_{\sigma} \rangle}{\langle n_b \rangle}$ increases for decreasing $\langle n_{\sigma} \rangle$. This behavior is illustrated in Fig.~3c where the dependence of $\phi_2$ on $\Delta \epsilon$ is displayed for two ensembles of CMC  events differing in the mean multiplicity of initial critical sigmas ($\langle n_{\sigma} \rangle \approx 31$ (full circles) and $\langle n_{\sigma} \rangle \approx 18$ (stars)). For both sets $\phi_2$ decreases with increasing $\Delta \epsilon$, however when $\langle n_{\sigma} \rangle$ is smaller the decrease is significantly slower. It must be noted that the quality of the power-law fits remains very good ($R^2 \approx 0.98$) even for $\Delta \epsilon \approx 50$~MeV and therefore the associated errors $\delta \phi_2$ are small. Finally, we have examined the influence of random pions not originating from the decay of critical sigmas on the obtained $\phi_2$ value. In Fig.~3d we present the results for $\phi_2$, calculated at threshold using the dipion invariant-mass window $[280,281]$~MeV, as a function of the percentage of random pions per event in the considered ensemble. We observe that the variation of $\phi_2$ is relatively slow and beyond 80\% $\phi_2$ drops rapidly to zero. In the same figure it is shown that up to 20\% background, the obtained critical index remains at the level of the QCD prediction. This last observation indicates also the limitations of the proposed subtraction method based on the correlator $\Delta F_2$ (eq.~\ref{eq:correlator}). Finally it is of interest to note that in the actual system of $\pi^+\pi^-$ pairs at threshold, weakly correlated pions of opposite charge originate mainly from the $I=2$ (isotensor) channel and form a negligible background of the sigma component \cite{Rapp03,CHAOS}. As a result we expect a small percentage of uncorrelated (weakly correlated) pions as we approach the critical point. This admixture, however, does not affect the extraction of the critical index $\phi_2$ from the experimental data (Fig.~3d).\\

\begin{figure}[htbp]
\centerline{\includegraphics[width=18 cm,height=14 cm]{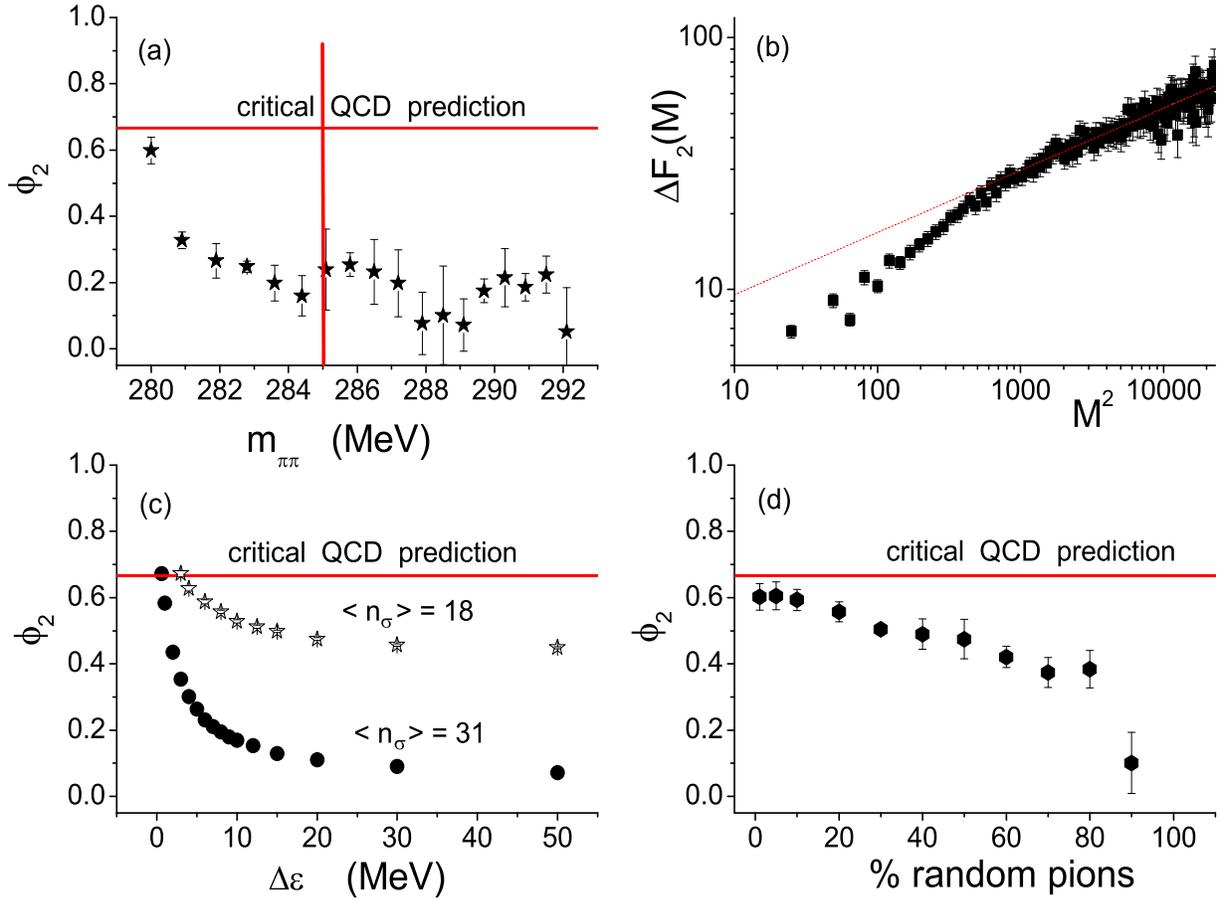}}
\caption{The results of an analysis of simulated CMC events are presented. In (a) we show the $\phi_2$ values calculated using non-overlapping dipion invariant-mass windows (with $\langle n_{\pi^+ \pi^-} \rangle \approx 2$) located at an increasing distance from the two-pion threshold. The vertical line at $285$~MeV indicates the upper limit of the dipion invariant mass region affected by Coulomb correlations. In (b) is shown the log-log plot of the correlator $\Delta F_2(M)$ for CMC events, calculated in the invariant-mass window $[285,286]$~MeV. The dashed line indicates the result of the linear fit in the range $M^2 \in [2000,22500]$. In (c) we show the dependence of $\phi_2$ on $\Delta \epsilon$ using for the analysis dipion invariant-mass windows $[280$~MeV,$280$~MeV $+ \Delta \epsilon]$ with $\Delta \epsilon$ increasing from $1$~MeV to $50$~MeV. Two different ensembles of CMC events with $\langle n_{\sigma} \rangle \approx 31$ (full circles) and $\langle n_{\sigma} \rangle \approx 18$ (stars) have been analyzed. Finally in (d) we plot the $\phi_2$ value at threshold calculated using an ensemble of 15000 CMC events contaminated with random pions, as a function of their percentage.}
\label{fig:fig3abcd}
\end{figure}

\vspace*{0.1cm}
\hspace*{0.3cm} In our calculations we have used an ensemble of 15000 CMC events with mean critical sigma multiplicity $\langle n_{\sigma} \rangle \approx 31$ leading after the decay to $\langle \pi^+ \rangle \approx 20$ with the exception of the analysis shown in Fig.~3c where we have used in addition an ensemble of 15000 CMC events with mean critical sigma multiplicity $\langle n_{\sigma} \rangle \approx 18$. A conclusion drawn from this study is that moving the $\pi^+\pi^-$-mass interval away from the mass of the critical sigma enhancement or increasing its size clearly decreases the fitted value of $\phi_2$. This behavior is attributed to the increasing contributions from combinatorial pair background. In addition the $\phi_2$-value decreases and the quality of the power-law fit is reduced, if the pion sector is highly contaminated by random pions.\\
\vspace*{0.1cm}

\hspace*{0.3cm} It is of interest to note here that, when the distance of the freeze-out state from the critical point increases, the spike in Fig.~2a is transformed to a smooth maximum located at an increasing distance away from the two-pion threshold \cite{Kunihiro99}. Our experience from the analysis of CMC data suggests that, in this case, the results of the fluctuation analysis in the reconstructed dipion sector should be characterized by three effects signaling the departure from the critical point:
\begin{itemize}
\item A decrease of the maximum $\phi_2$ value,
\item a displacement of the location of this maximum to invariant-mass values greater than the two-pion threshold and
\item a reduction of the quality of the power-law fits of $\Delta F_2$ (larger $\chi^2 /dof$, smaller $R^2$).
\end{itemize}
\vspace*{0.1cm}

\hspace*{0.3cm} The CMC data do not incorporate Coulomb correlations between charged pions and therefore no lower limit in the value of $\epsilon_1$ in (\ref{eq:epswin}) is necessary. However, in the analysis of a real system one has to take the constraint $\epsilon_1 \geq 5$~MeV into account. Based on the results of the sigma reconstruction in the CMC events (Fig.~3a), we expect that the signature of the critical sigma correlations is a global maximum in $\phi_2$ as a function of $m_{\pi^+\pi^-}$ the location of which tends towards
$2 m_{\pi}$ with its value approaching the QCD prediction $\phi_2=2/3$. In the search for this maximum one has also to verify the quality of the power-law behavior.  Having as a guide the experience gained by the reconstruction of the critical sigma sector in the CMC events, we will present in the next section the results of a similar analysis applied to the four SPS data sets described previously.\\

\section[roman]{Results of the data analysis at the SPS}

\hspace*{0.3cm} The algorithm described in section III can be directly applied to the A+A data sets (A = p, C, Si, Pb) measured at 158$A$~GeV. The results of this analysis are presented step-by-step in the following. For each system $A$ we first
determine the size $\Delta \epsilon$ of the kinematical window in (\ref{eq:epswin}) in order to achieve an optimal mean multiplicity of dipions in the considered domain allowing at the same time for sufficient statistics. It turns out that
$\langle n_{\pi^+\pi^-} \rangle_{\Delta \epsilon} \approx 4$ is a good choice leading to small statistical fluctuations in the calculated $\phi_2$ values for the C+C and Si+Si systems. To achieve this multiplicity in the p+p system one has to choose a large $m_{\pi\pi}$ mass window. On the other hand, to obtain this multiplicity value in the Pb+Pb system one has to decrease $\Delta \epsilon$ below the experimental invariant-mass resolution $\delta \epsilon$ of the NA49 detector. This is seen in Fig.~4 where the functions $\Delta \epsilon(m_{\pi^+\pi^-})$ for $\langle n_{\pi^+\pi^-} \rangle_{\Delta \epsilon} = 4$ are presented for the systems C+C (crosses), Si+Si (full circles) and Pb+Pb (open triangles). In the same plot the solid line displays the experimental invariant-mass resolution $\delta \epsilon$ as a function of the dipion invariant mass $m_{\pi^+\pi^-}$ assuming constant momentum transfer resolution $\delta Q \approx 5$~MeV (where $Q=\sqrt{-(p_{\pi^+}-p_{\pi^-})^2}$ for a pair of oppositely charged pions).\\

\begin{figure}[htbp]
\centerline{\includegraphics[width=8.5 cm,height=10.5 cm]{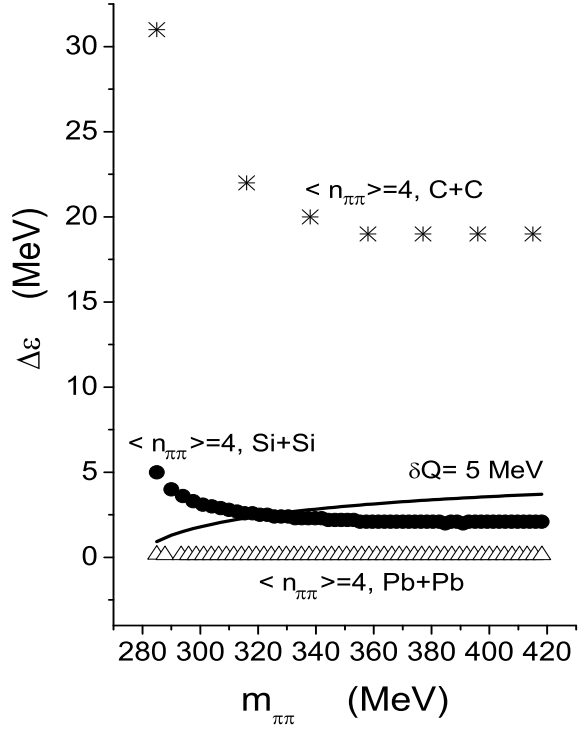}}
\caption{The functions $\Delta \epsilon(m_{\pi\pi})$ corresponding to $\langle n_{\pi^+\pi^-} \rangle_{\Delta \epsilon} \approx 4$
for the systems C+C (crosses), Si+Si (full circles) and Pb+Pb (open triangles). For comparison the line $\delta \epsilon(m_{\pi\pi})$ corresponding to the NA49 experimental resolution of the momentum transfer $\delta Q \approx 5$~MeV is also shown.}
\label{fig:fig4}
\end{figure}

\vspace*{0.1cm}
\hspace*{0.3cm} It is obvious that the analysis is meaningful only as long as $\Delta \epsilon(m_{\pi\pi})$ is larger than $\delta \epsilon(m_{\pi\pi})$. This constraint does not affect the C+C system, while it restricts the region of analysis for Si+Si to the domain
$[285, 320]$~MeV. For the Pb+Pb system it is impossible to satisfy the constraint $\Delta \epsilon > \delta \epsilon$ for $\langle n_{\pi^+\pi^-} \rangle_{\Delta \epsilon} \approx 4$. Thus, in this case one can only use $\Delta \epsilon$ values which are larger than $\delta \epsilon$.\\
\vspace*{0.1cm}

\hspace*{0.3cm} The fluctuation analysis of the four SPS systems can be summarized as follows:
\begin{itemize}
\item{For the p+p system we have considered a single $m_{\pi\pi}$ mass window with $\epsilon_1=5$~MeV and $\Delta \epsilon=290$~MeV (much larger than $\delta \epsilon$) leading to $\langle n_{\pi^+\pi^-} \rangle_{\Delta \epsilon} \approx 4$. The index $\phi_2$ is found to be close to zero with
$R^2 \approx 0$ indicating the absence of power-law sigma correlations in the considered kinematical region.}
\item{In C+C data the region $[285,350]$~MeV has been covered using non-overlapping mass intervals of varying size in order to achieve
$\langle n_{\pi^+\pi^-} \rangle_{\Delta \epsilon} \approx 4$ within each interval and the corresponding $\phi_2$ is calculated. In order to scan completely the considered kinematical region we have used three different sets of non-overlapping invariant-mass intervals differing in the starting $\epsilon_1$ value: $\epsilon_1=5$~MeV (set 1), $\epsilon_1=15$~MeV (set 2) and $\epsilon_1=25$~MeV (set 3). It turns out that also in the C+C system the calculated index $\phi_2$ is around zero becoming also slightly negative in some invariant-mass intervals (see Fig.~6a). The dependence on the location of the invariant-mass window used in the analysis is weak and the quality of the power-law fits poor ($R^2 \to 0$, see Fig.~6c). No obvious maximum of $\phi_2$ occurs in the considered region.  In fact for C+C it is possible to perform the intermittency analysis using smaller
invariant-mass intervals in order to approach the condition of optimal multiplicity $\langle n_{\pi^+\pi^-} \rangle_{\Delta \epsilon} \approx 2$ for the suppression of the combinatorial background \cite{Ant05}. We have performed also this analysis in the region $[285,320]$~MeV and the results remained practically unchanged. The largest value of $\phi_2$ was found to be $\phi_2=0.02 \pm 0.02$ in the interval $[285,290]$~MeV with $R^2 \approx 0.01$. Thus, the C+C system shows no signature of critical transverse momentum fluctuations in the sigma-sector according to our analysis.}

\item{For the Si+Si system the dipion invariant-mass region ($[285,320]$~MeV) is covered using non-overlapping intervals of varying size, as in the case of C+C, in order to achieve $\langle n_{\pi^+\pi^-} \rangle_{\Delta \epsilon} \approx 4$. Similarly to the C+C case we use three different sets of invariant-mass intervals differing in the initial $\epsilon_1$ value in order to achieve a complete scan of the aforementioned kinematical region: $\epsilon_1=5$~MeV (set 1), $\epsilon_1=7$~MeV (set 2) and $\epsilon_1=8.5$~MeV (set 3). A clear maximum of $\phi_2$ at
$m_{\pi^+\pi^-} \approx 302$~MeV is here observed (see Fig.~6b,d) leading to $\phi_{2,max} \approx 0.33 \pm 0.04 $ ($\chi^2 /dof \approx 0.3$ and $R^2 \approx 0.71$). Thus, Si+Si shows properties characterizing a system freezing out at a relatively small distance from the critical point such that remnants of critical fluctuations are present in the transverse momenta of the produced pions. In fact the location of the maximum is close to the two-pion threshold, its value $\phi_{2,max}$ is large and the quality of the corresponding power-law fit, measured through $R^2$, is good.}

\item{Finally, the analysis of the Pb+Pb system has been restricted to the interval $[285,286]$~MeV using $\Delta \epsilon=1$~MeV for $\epsilon_1=5$~MeV leading to $\langle n_{\pi^+\pi^-} \rangle_{\Delta \epsilon} \approx 20$. The corresponding experimental resolution of the dipion invariant mass in this region is $\delta \epsilon \approx 0.93$~MeV. The critical index $\phi_2$ is found vanishingly small ($\phi_2 \approx 0.04$) with
a fit error of $\delta \phi_2 \approx 0.02$, and $R^2 \approx 0.07$. In order to clarify the role of high multiplicity near threshold we have performed a CMC simulation using an ensemble of 1500 events with $\langle n_{\sigma} \rangle \approx 250$ in two intervals of the invariant mass:\\

\noindent
(a) at the $2 m_{\pi}$ threshold where the singularity of the $\sigma$-enhancement prevails, taking $\langle n_{\pi^+\pi^-} \rangle \approx 29$ in the $m_{\pi^+\pi^-}$ window $[280,280.6]$~MeV. We found a pattern close to the critical one with $\phi_2=0.47 \pm 0.01$, $R^2=1.00$  which is robust against changes of the average multiplicity $\langle n_{\pi^+\pi^-} \rangle$ and \\

\noindent
(b) in the nearby $m_{\pi^+\pi^-}$ interval $[285,285.3]$~MeV where the actual analysis of the experimental data for Pb+Pb was performed, taking $\langle n_{\pi^+\pi^-} \rangle_{\Delta \epsilon} \approx 38$. We found a non-critical pattern with $\phi_2 \approx 0.07 \pm 0.01$, comparable to the value measured in the above analysis for Pb+Pb. This drastic change in the CMC result is due to the high multiplicity of non-critical $\pi^+\pi^-$pairs in windows of invariant mass at a distance from the singularity ($\sigma$-enhancement) at threshold. However, the correlator $\Delta F_2$ of the CMC system remains close to a power-law ($R^2 \approx 0.93$) in contrast to the situation in Pb+Pb ($R^2 \approx 0$). The conclusion drawn from this study using CMC events is that in large systems the reconstruction of critical fluctuations in the dipion sector seems not possible with our method.}
\end{itemize}
\vspace*{0.1cm}

\hspace*{0.3cm} In Figs. 5(a-d) we show the dipion ($\pi^+\pi^-$ pairs) invariant-mass distributions with a characteristic smooth maximum for all A+A systems we have analyzed. The maximum of the distribution for C+C is located at $m_{\pi^+\pi^-} \approx 421$~MeV while for Si+Si it lies at
$m_{\pi^+\pi^-} \approx 386$~MeV. Apart from this slight displacement of the maximum of the dipion invariant-mass distribution towards the two-pion threshold for Si+Si, the plots are similar to that of Fig.~2b obtained from CMC events. As expected, the reconstructed invariant-mass distribution alone cannot reveal the underlying activity of critical sigmas in the freeze-out state. Obviously the study of fluctuations is a necessary tool for this purpose.\\

\begin{figure}[htbp]
\centerline{\includegraphics[width=18 cm,height=14 cm]{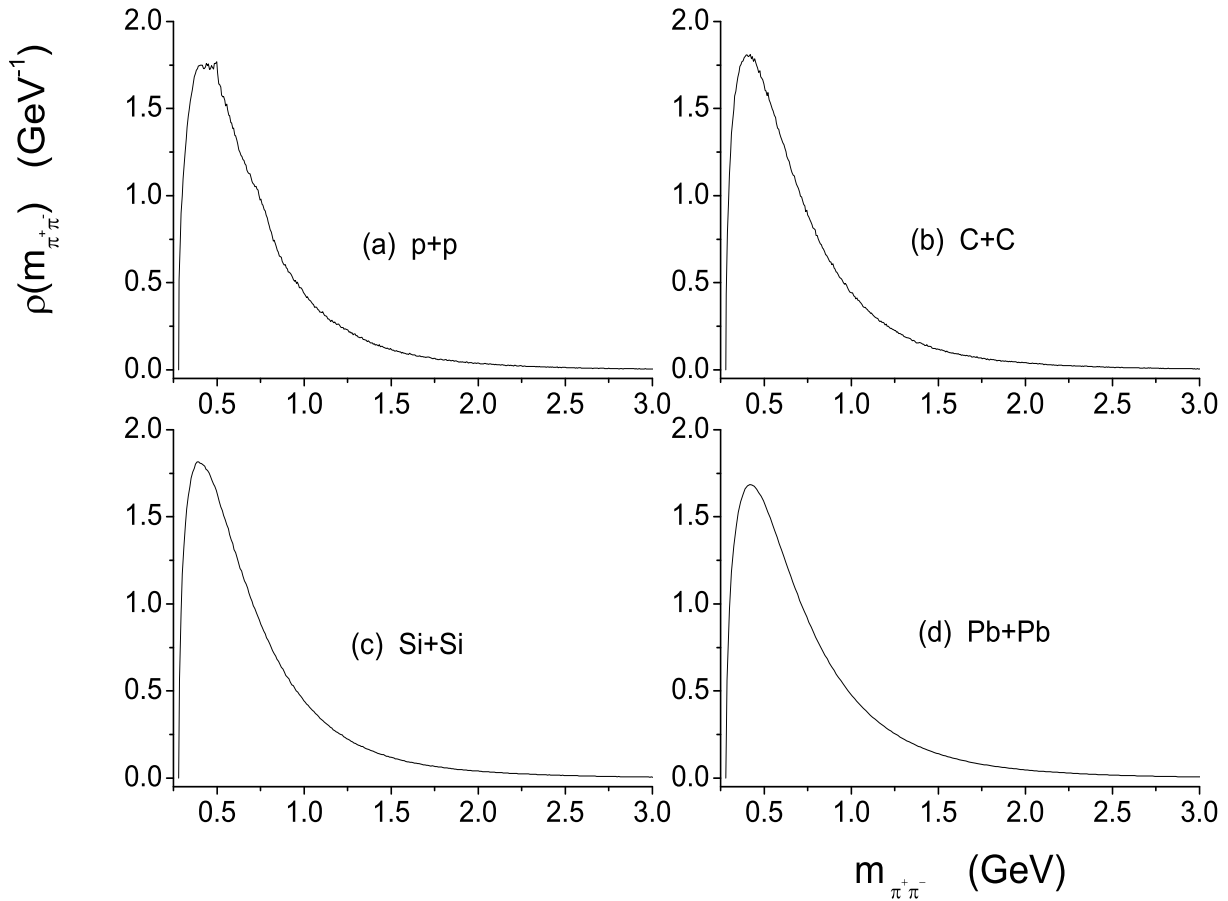}}
\caption{The invariant-mass distribution for $\pi^+\pi^-$ pairs obtained from the (a) p+p, (b) C+C, (c) Si+Si and (d) Pb+Pb NA49 data at maximum SPS energy.}
\label{fig:fig5abcd}
\end{figure}

\vspace*{0.1cm}
\hspace*{0.3cm} The detailed results of the search for critical fluctuations in the
C+C and Si+Si systems are presented in Figs.~6(a-d) and 7(a-f). In particular  Fig.~6a shows $\phi_2$ for C+C in the kinematical range $[285,350]$~MeV while Fig.~6b shows the corresponding plot for Si+Si in the kinematical range $[285,320]$~MeV. One can clearly see the absence of power-law fluctuations in C+C for the entire range of analysis and the formation of a pronounced maximum with $\phi_{2,max} \approx 0.35$ located at
$m_{\pi^+\pi^-} \approx 302$~MeV in Si+Si. The quality of the power-law fits in this analysis is presented in Figs.~6c (for C+C system) and 6d (for Si+Si system) where we plot the corresponding $R^2$-values. It is seen that in the C+C system there is no $m_{\pi^+\pi^-}$ region where the correlator $\Delta F_2(M)$ is close to a power-law since all the $R^2$-values are found to be small. On the other hand, in the Si+Si system there is a peak in $R^2$ at the same position as in $\phi_2$ ($m_{\pi^+\pi^-} \approx 302$~MeV) with a value ($R^2_{max} \approx 0.7$) indicating that in this
invariant-mass window the correlator is well described by a power-law. The form of the factorial moments in transverse momentum space for Si+Si both for the real data as well as for the mixed events at the maximum and at a distance (below or above) from it are shown in the left column of Fig.~7. The corresponding correlator $\Delta F_2$ is presented in the right column. In the region of the maximum one can clearly distinguish $F_2$ of the data from $F_2$ of the mixed events. However, when the interval of analysis lies at a distance from the maximum then the values of $F_2$ for the data overlap with those of $F_2$ for mixed events. As a consequence the calculated value of $\phi_2$ decreases and the quality of the power-law fits is significantly reduced. A possible dependence of the obtained results for $\phi_2$ on the set of mixed events used for the background subtraction is suppressed by averaging in all the calculations of $\Delta F_2$ over $10$ sets of mixed events generated using different sequences of random numbers. In order to avoid systematic biases adequate 2-track and momentum resolution are required. The effects of momentum resolution were investigated by smearing the measured track momenta by the upper limit of the transverse momentum uncertainty of $\frac{\Delta p}{p^2} \approx 7 \cdot 10^{-4}$ (GeV/c)$^{-1}$ and then recalculating the correlator $\Delta F_2$ in the mass window $[302.1,305.1]$~MeV. No significant change of
either $\Delta F_2$ or $\phi_2$ was observed. Possible effects of limited 2-track resolution were investigated as follows. Distributions
of average track multiplicities (cell occupancies) in dependence of the cell number $M^2$ were computed for both real and mixed events. The ratio
of these distributions was found to be independent of $M^2$. In particular there was no decrease at the largest $M^2$ (smallest phase
space cells) used in the analysis demonstrating fully sufficient 2-track resolution.\\

\vspace*{0.1cm}
\hspace*{0.3cm} In order to test the validity of the error attached by the fit to the value of $\phi_2$ we subdivided the C+C and Si+Si data each into 4 subsets with almost equal number of events. As uncertainty estimate of $\phi_2$ for the full data samples we take half the spread of the $\phi_2$ values determined for the 4 independent sub-samples. It turns out that this estimate is close to the error given by the fit for $\phi_2$ from the full data samples. We observe that the worsening of the statistics has impact on the quality of the power-law fits but retains the signature of the maximum in the Si+Si system.\\

\begin{figure}[htbp]
\centerline{\includegraphics[width=20 cm,height=14 cm]{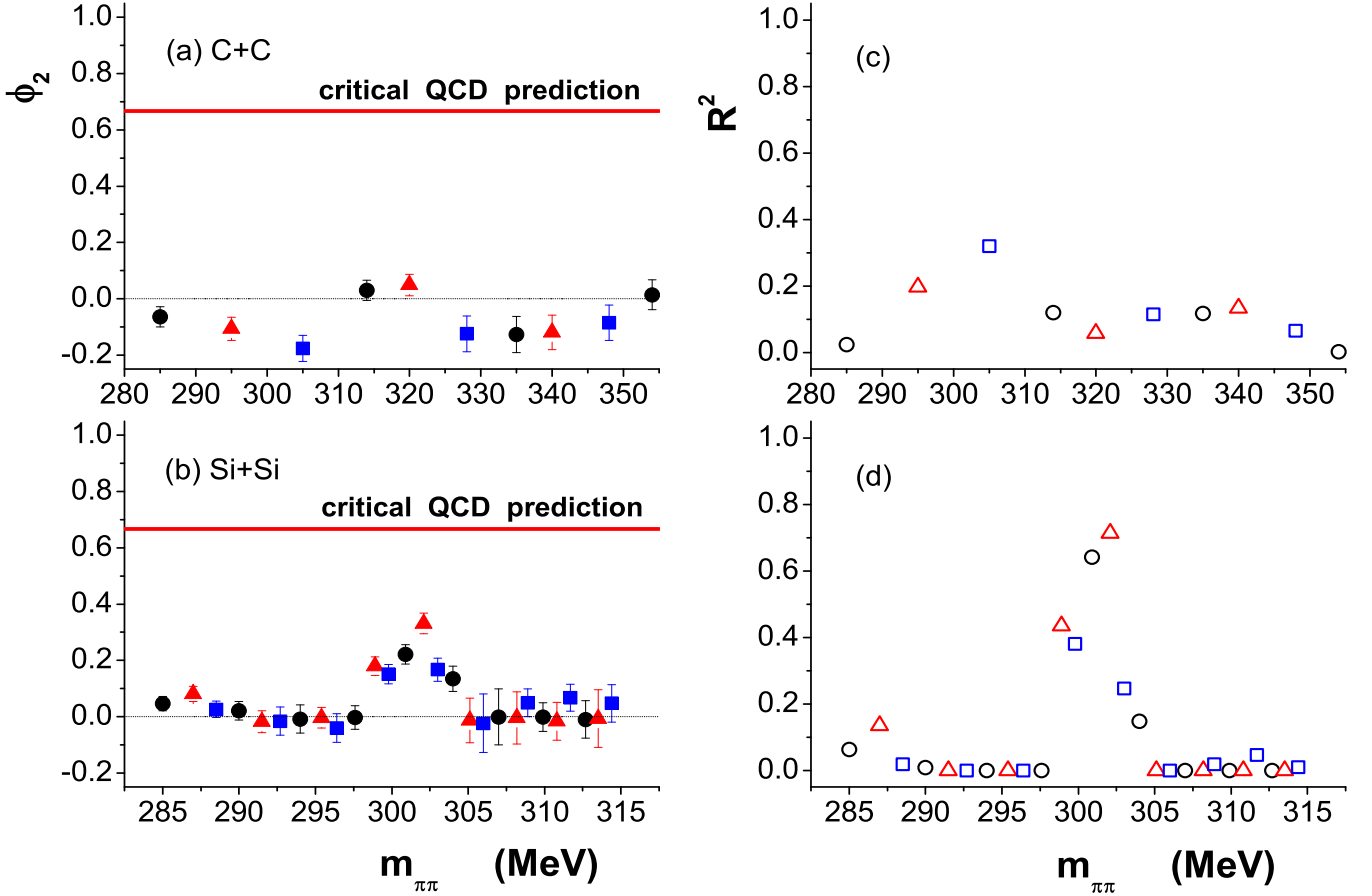}}
\caption{The function $\phi_2(m_{\pi\pi})$ (a) for the C+C system and (b) for the Si+Si system. The $R^2$-values of the corresponding power-law fits are displayed in (c) and (d) respectively. The errors here are determined by the fit. The different symbols are used to indicate the three different sets of dipion invariant-mass intervals used in the analysis as described in the text.}
\label{fig:fig6abcd}
\end{figure}

\begin{figure}[htbp]
\centerline{\includegraphics[width=16 cm,height=20 cm]{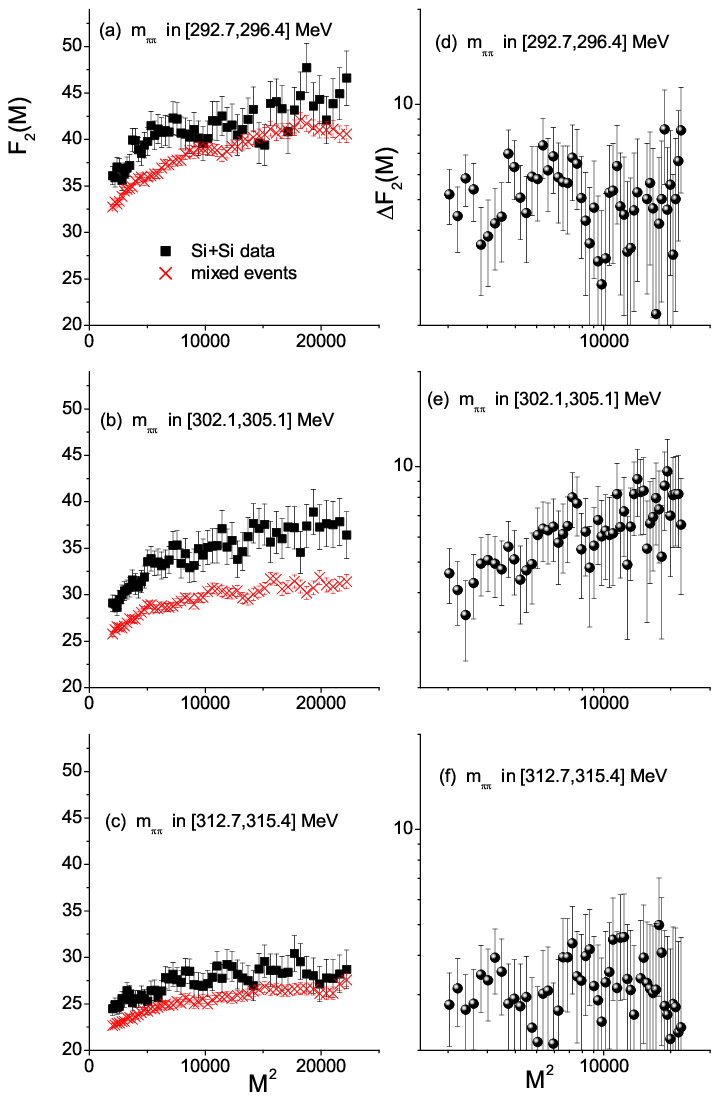}}
\caption{The second factorial moment $F_2$ for Si+Si data (solid squares) and the corresponding mixed events (crosses) obtained by intermittency analysis in three different invariant-mass intervals: (a) $[296.4,297.2]$~MeV, (b) $[302.1,305.1]$~MeV and (c) $[312.7,315.4]$~MeV. The corresponding subtracted moments $\Delta F_2$ are shown in (d), (e) and (f), respectively.}
\label{fig:fig7abcdef}
\end{figure}

\vspace*{0.1cm}
\hspace*{0.3cm} For a complete presentation of the performed reconstruction analysis we show in Fig.~8(a-d) the factorial moments of the four considered SPS systems. For Si+Si we show the moment (full triangles) calculated in the dipion invariant-mass window for which  $\phi_2$ is maximized while for the C+C system we show the moment for the invariant mass window located as close as possible to the two-pion threshold ($m_{\pi^+\pi^-} \in [285,314]$~MeV). In the same plot we also display the second moments for the corresponding mixed events (open triangles). In addition in Fig.~9(a-d) we give the subtracted moments $\Delta F_2$.\\

\begin{figure}[htbp]
\centerline{\includegraphics[width=12 cm,height=16 cm]{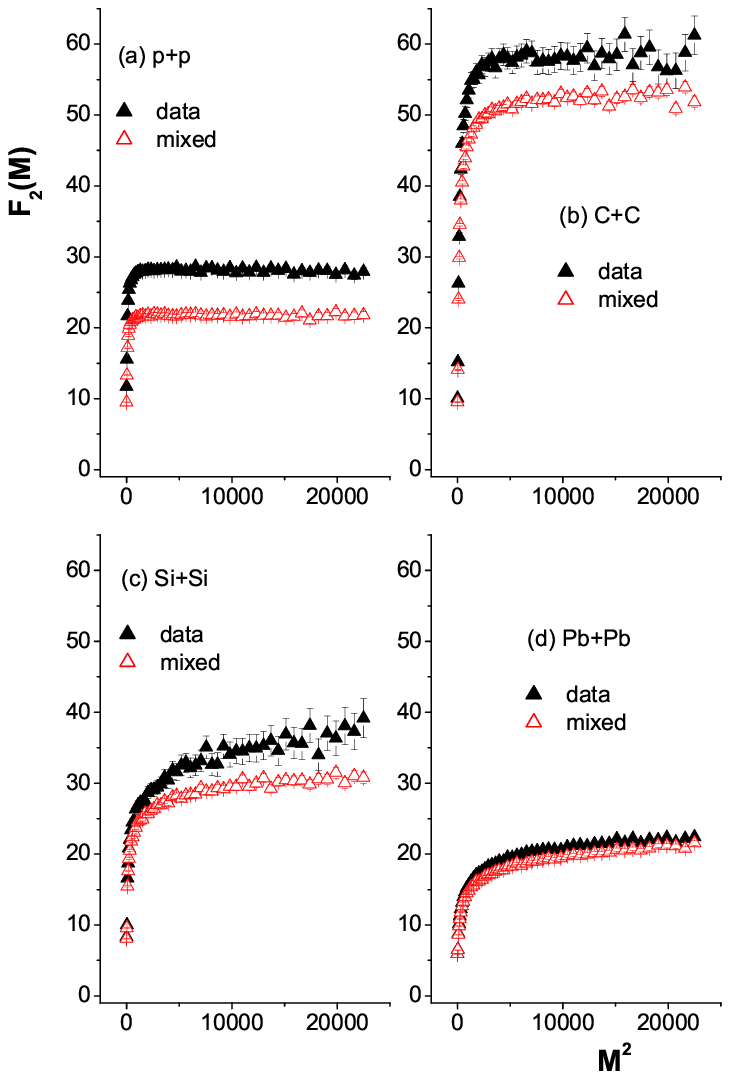}}
\caption{The second factorial moment in transverse momentum space for: (a) p+p (window of analysis $[280,570]$~MeV), (b) C+C (window of analysis $[285,314]$~MeV), (c) Si+Si (window of analysis $[300.9,304]$~MeV) and (d) Pb+Pb (window of analysis $[285,286]$~MeV) systems. The full triangles represent the moments of NA49 data while the open triangles the moments for the corresponding mixed events.}
\label{fig:fig8abcd}
\end{figure}

\begin{figure}[htbp]
\centerline{\includegraphics[width=18 cm,height=12 cm]{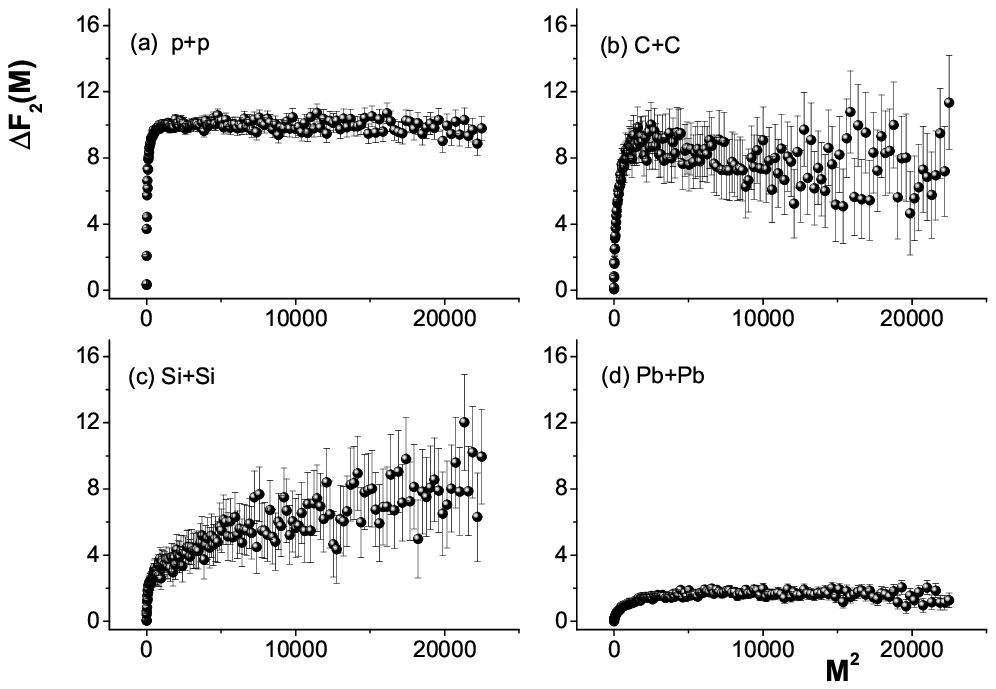}}
\caption{The combinatorial background subtracted moments $\Delta F_2$ in transverse momentum space for: (a) p+p, (b) C+C, (c) Si+Si and (d) Pb+Pb systems.}
\label{fig:fig9abcd}
\end{figure}

\vspace*{0.1cm}
\hspace*{0.3cm} In Fig.~10 the dependence of $\phi_2$ on the size $A$ of the considered system is illustrated. The horizontal straight line at
$\phi_2=2/3$ is drawn to indicate the critical QCD prediction \cite{Ant05}. In addition we show the line $\phi_2=0$ to guide the eye. The shaded region in Fig.~10 indicates the $A$-values for which the algorithm of reconstruction described in section III fails to reveal, even partially, existing critical fluctuations without entering into the Coulomb region ($m_{\pi^+\pi^-} \leq 285$~MeV). The limiting value $A=82$ is determined using CMC simulated events. As previously described, the errors are determined by the spread of the $\phi_2$ values calculated after subdivision of the entire event ensemble in four equally large subsets. We do not include in the plot the result for the Pb+Pb system since $A > 82$ and also because the actual measurements, in this case, cannot fulfill the constraint of mean dipion multiplicity, $\langle n_{\pi^+\pi^-} \rangle \approx 4$.\\

\begin{figure}[htbp]
\centerline{\includegraphics[width=15 cm,height=12 cm]{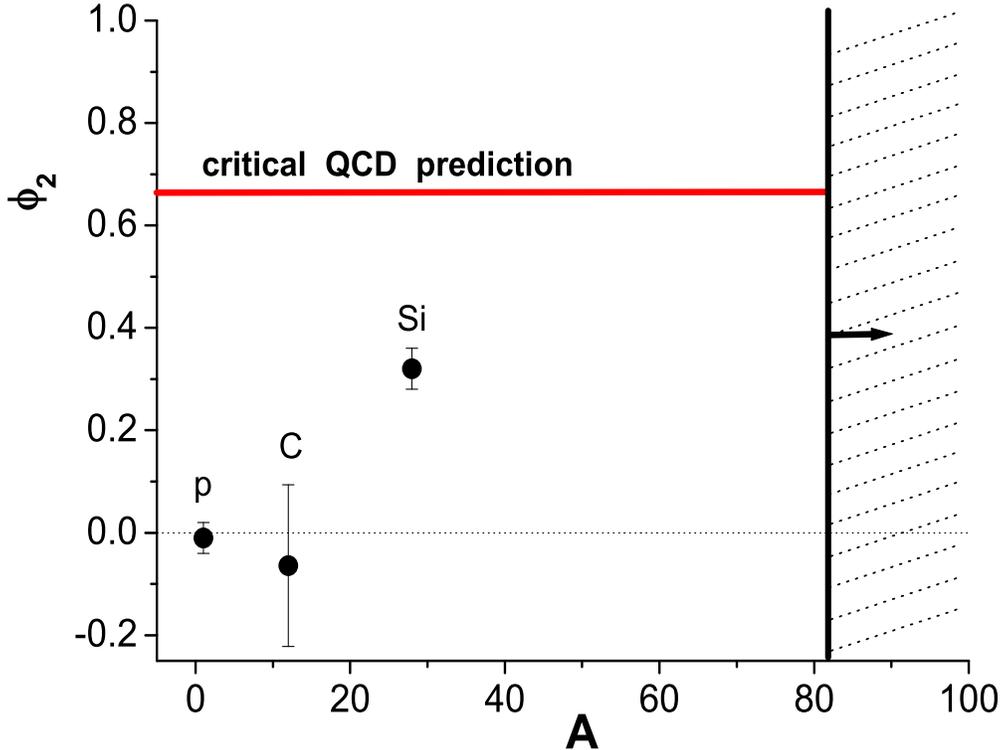}}
\caption{The fitted values of $\phi_2$ for the A+A systems (A = p, C, Si) studied by NA49 as a function of the size A. All three systems obey  $\langle n_{\pi^+\pi^-} \rangle \approx 4$ in the corresponding window of analysis. The upper horizontal line presents the theoretically expected value ($2/3$) for a system freezing out at the QCD critical point while the lower horizontal line is at $\phi_2=0$. The shaded region indicates the $A$-values for which the reconstruction algorithm of section III is not conclusive. The shown error bars have been obtained by analyzing sub-samples (see text).}
\label{fig:fig10}
\end{figure}

\vspace*{0.1cm}
\hspace*{0.3cm} Finally in Figs.~11a,b,c we present a comparison of $\Delta F_2$ for Si+Si (a) with the corresponding moments in CMC (b) and HIJING (c). The analysis for HIJING was made using the same detector acceptance and $m_{\pi\pi}$ mass windows as employed for the NA49 data. The CMC result is obtained from the reconstruction analysis in the dipion invariant-mass window $[280,280.6]$~MeV leading to a mean dipion multiplicity of 1.6 and optimal reconstruction of critical fluctuations ($\phi_2=0.67 \pm 0.01$ and $R^2=1$). The quality of the power-law fit and the slope for the NA49 data show a similar behavior with the CMC events and deviate significantly from the behavior found for the HIJING data where no reasonable power-law fit gets possible since the corresponding $R^2$ value tends to zero ($\phi_2=0.02 \pm 0.09$ with $R^2=0.002$). Contrary to Si+Si, the fluctuations in the C+C NA49 system, as shown in Figs.~11d,e, are comparable with those in the HIJING events ($\phi_2=-0.003 \pm 0.03$ with $R^2=0.01$).\\

\begin{figure}[htbp]
\centerline{\includegraphics[width=20 cm,height=15 cm]{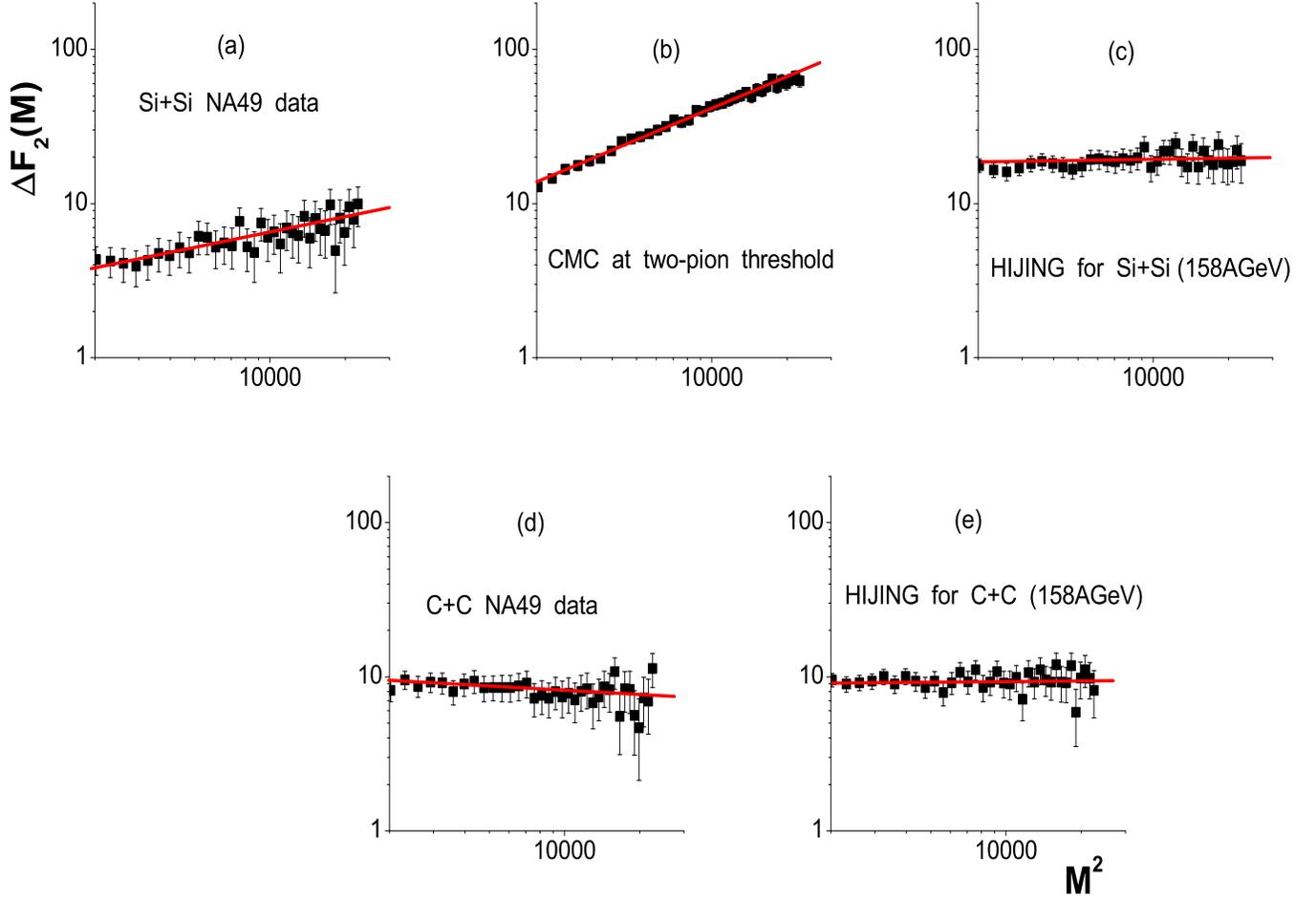}}
\caption{The correlator $\Delta F_2$ corresponding to the best solution for Si+Si at 158$A$~GeV using:
(a) NA49 data, (b) CMC generated events and (c) HIJING events. The dotted lines display the corresponding linear fit in log-log scale. For comparison we also show representative results for C+C at 158$A$~GeV using: (d) NA49 data and (e) HIJING events.}
\label{fig:fig11}
\end{figure}

\section[roman]{Discussion and Conclusions}

\hspace*{0.3cm} In this work the behavior of pion pairs $(\pi^+\pi^-)$ produced near the two-pion threshold in nuclear collisions at the CERN SPS, is investigated. The motivation for such a study originates from the fact that the above system $(\pi^+\pi^-,~m_{\pi^+\pi^-}~\stackrel{>}{\sim}~2 m_{\pi})$ has a
strong component in the $\s$-mode \cite{Kunihiro99,CHAOS} (scalar and isoscalar) and therefore it is sensitive to the order parameter of the QCD critical point \cite{Wilcz00}. As a consequence, it develops unconventional density fluctuations with a power-law behavior, characteristic of a second-order phase transition, provided that the particle system produced in these collisions freezes out close to the critical endpoint in the QCD phase diagram \cite{Wilcz00,Ant05}.
Using critical events generated by the CMC algorithm, it is shown that a global maximum of the suitably defined critical index $\phi_2$ appears at a value approaching the QCD prediction $(\phi_2 \approx 2/3)$ when moving towards the two-pion threshold. This behavior constitutes a signature for the existence of the critical point and the partial restoration of chiral symmetry.\\

\vspace*{0.1cm}
\hspace*{0.3cm} To search for such a signature we have analyzed $\pi^+\pi^-$ pairs from p+p, C+C, Si+Si and Pb+Pb collisions at the maximum SPS energy of 158$A$~GeV. We have chosen scaled factorial moments of second order in the transverse momentum plane and in the $\sigma$-mode as the basic observables in which the power-laws of critical QCD fluctuations can be revealed. In particular the exponent $\phi_2$ of the power-law behavior in small domains of the momentum space (intermittency) provides us with a signature of critical fluctuations when compared to the QCD value: $\phi_2=2/3$. The results of our analysis are summarized as follows:
\begin{itemize}
\item{Large power-law fluctuations, measured through the index $\phi_2$ of factorial moments are developed in the system Si+Si at 158$A$~GeV ($\phi_2 \approx 0.35$) .}
\item{The observed fluctuations may have unconventional origin as suggested by comparison with the corresponding moments in HIJING and CMC.}
\item{In Pb+Pb collisions, the high multiplicity of the produced pions combined with the restrictions imposed by the necessity to exclude the Coulomb correlations and the resolution of the experiment, decrease the sensitivity to the sigma fluctuations near the two-pion threshold. Whether the vanishingly small value of $\phi_2$ found in this case is due to the effect of high multiplicity or to a genuine non-critical nature of the freeze-out state of the system, cannot be resolved without penetrating the Coulomb region to reach the $2 m_{\pi}$ threshold.}
\end{itemize}

\vspace*{0.1cm}
\hspace*{0.3cm} In conclusion, a sizable effect of $\pi^+\pi^-$ pair fluctuations with critical characteristics was found in Si+Si collisions at 158$A$~GeV
(the chemical freeze-out parameters extracted using the hadron gas model \cite{Becat04} for this reaction are: $\mu_B \approx 253$~MeV,
$T \approx 163$~MeV). This effect may be associated with the presence of the QCD critical point in the wider SPS region. Complementary studies in the baryonic sector are necessary in order to clarify the picture \cite{Ant06} and overcome the limitations and uncertainties still remaining in the reconstruction of the critical $\sigma$-mode.\\

\vspace*{0.1cm}
\hspace*{0.3cm} The overall outcome of this investigation combined with theoretical estimates \cite{Fodor04}, based in particular on Lattice QCD at high temperature, suggest that an intensive experimental search for the QCD critical point in the region of the phase diagram $180$~MeV $\leq \mu_B \leq 400$~MeV, $150$~MeV $\leq T \leq 170$~MeV is of high interest. Such a program is proceeding both at the CERN SPS \cite{LoI} and BNL RHIC
\cite{RHICW} and will hopefully strengthen the evidence for the critical point. \\
\vspace*{0.1cm}

\noindent
{\bf{Acknowledgements}} This work was supported by the US Department of Energy Grant DE-FG03-97ER41020/A000, the Bundesministerium fur Bildung und Forschung, Germany (06F137), the Virtual Institute VI-146 of Helmholtz Gemeinschaft, Germany, the Polish State Committee for Scientific Research     (1 P03B 006 30, 1 P03B 097 29, 1 PO3B 121 29, 1 P03B 127 30), the Hungarian Scientific Research Foundation (T032648, T032293, T043514), the Hungarian National Science Foundation, OTKA, (F034707), the Polish-German Foundation, the Bulgarian National Science Fund (Ph-09/05), the Croatian Ministry of Science, Education and Sport (Project 098-0982887-2878), Stichting FOM, the Netherlands. Partial financial support through the research programs ``Pythagoras'' of the EPEAEK II (European Union and the Greek Ministry of Education) and ``Kapodistrias'' of the University of Athens are also acknowledged.


\vspace*{0.1cm}

\begin{thebibliography}{999}

\bibitem{Wilcz00} M. Asakawa and K. Yazaki, Nucl. Phys. {\bf A504}, 668 (1989); A. Barducci, R. Casalbuoni,
S. De Curtis, R. Gatto and G. Pettini,
Phys. Lett. {\bf B231}, 463 (1989); F. Wilczek, hep-ph/0003183; M. A. Halasz, A. D. Jackson, R. E. Shrock, M. A. Stephanov and J. J. M. Verbaarschot,
Phys. Rev. {\bf D58}, 096007 (1998); M. Stephanov, K. Rajagopal, E. Shuryak, Phys. Rev. Lett. {\bf 81}, 4816 (1998);
J. Berges, N. Tetradis and C. Wetterich, Phys. Rep. {\bf 363}, 223 (2002); M. A. Stephanov, Prog. Theor. Phys. Suppl. {\bf 153}, 139 (2004); Int. J. Mod. Phys. {\bf A20}, 4387 (2005); hep-ph/0402115; R. Casalbuoni, POS CPOD2006:001 (2006); hep-ph/0610179.

\bibitem{Ant05} N. G. Antoniou, Y. F. Contoyiannis, F. K. Diakonos, A. I. Karanikas, C. N. Ktorides, Nucl. Phys. {\bf A693}, 799 (2001);
N. G. Antoniou, Y. F. Contoyiannis, F. K. Diakonos, G. Mavromanolakis, Nucl. Phys. {\bf A761}, 149 (2005).

\bibitem{Stan87} H. E. Stanley: {\it Introduction to Phase Transitions and Critical Phenomena}, Oxford University Press, 1987;
P. M. Chaikin, T. C. Lubensky: {\it Principles of condensed matter physics}, Cambridge University Press, 1997.

\bibitem{Kunihiro99} T. Hatsuda, T. Kunihiro and H. Shimizu, Phys. Rev. Lett. {\bf 82},
2840 (1999); T. Hatsuda and T. Kunihiro, Phys. Rep. {\bf 247}, 221 (1994); S. Chiku and T. Hatsuda,
Phys. Rev. {\bf D58}, 076001 (1998).

\bibitem{CHAOS} CHAOS Collaboration, N. Grion {\it et al.}, Nucl. Phys. {\bf A763}, 80 (2005).

\bibitem{Fodor04} Z. Fodor, S. D. Katz, J. High Energy Phys. {\bf 0404}, 050 (2004); {\bf 0203}, 014 (2002);
C. R. Allton, S. Ejiri, S. J. Hands, O. Kaczmarek, F. Karsch, E. Laermann, C. Schmidt,
Nucl. Phys. Proc. Suppl. {\bf B129}, 614 (2004); Phys. Rev. {\bf D68}, 014507 (2003);
R. V. Gavai, S. Gupta, Phys. Rev. {\bf D71}, 114014 (2005);
N. G. Antoniou, A. S. Kapoyannis, Phys. Lett. {\bf B563}, 165 (2003);
N. G. Antoniou, F. K. Diakonos, A. S. Kapoyannis, Nucl. Phys. {\bf A759}, 417 (2005);
M. I. Gorenstein, M. Ga\'{z}dzicki, W. Greiner, Phys. Rev. {\bf C72}, 024909 (2005);
N. G. Antoniou, F. K. Diakonos, A. S. Kapoyannis, Phys. Rev. {\bf C81}, 011901(R) (2010).

\bibitem{NA49041} NA49 Collaboration, S. V. Afanasiev {\it et al.}, Phys. Rev. {\bf C66}, 054902 (2002);
NA49 Collaboration, M. Ga\'{z}dzicki {\it et al.}, J. Phys. {\bf G30}, S701 (2004); R. Stock, hep-ph/0404125;
NA49 Collaboration, C. Alt {\it et al.}, Phys. Rev. {\bf C78}, 034914 (2008).

\bibitem{NA49042} M. Ga\'{z}dzicki and M. I. Gorenstein, Acta Phys. Polon. {\bf B30}, 2705 (1999)
[arXiv:hep-ph/9803462]; M. I. Gorenstein, M. Ga\'{z}dzicki and K. A. Bugaev, Phys. Lett. {\bf B567}, 175 (2003)
[arXiv:hep-ph/0303041]; M. Bleicher, hep-ph/0509314.

%\bibitem{Wolf96} E. A. De Wolf, I. M. Dremin, W. Kittel, Phys. Rept. {\bf 270}, 1 (1996).

\bibitem{Bial86} A. Bialas and R. Peschanski, Nucl. Phys. {\bf B273}, 703 (1986);
Nucl. Phys. {\bf B308}, 857 (1988).

\bibitem{Afan99} NA49 Collaboration, S. V. Afanasiev {\it et al.}, Nucl. Instrum.
Meth. {\bf A430}, 210 (1999).

\bibitem{Appel98} NA49 Collaboration, H. Appelshauser {\it et al.}, Eur. Phys. J.
{\bf A2}, 383 (1998).

\bibitem{Sey98} NA49 Collaboration, P. Seyboth {\it et al.}, Proceedings of the 8th
International Workshop on Multiparticle Production:
{\it "Correlations and fluctuations '98"}, Matrahaza, World Scientific (1999), pg.168.

\bibitem{Eve02} B. S. Everitt,{\it "The Cambridge Dictionary of Statistics"}, Cambridge University Press, 2002.

\bibitem{deWolf96} E. A. De Wolf, I. M. Dremin and W. Kittel, Phys. Rep. {\bf 270}, 1 (1996);
M. I. Adamovich {\it et al.} (EMU-01), Phys. Lett. {\bf B263}, 539 (1991); Nucl. Phys. {\bf B388}, 3 (1992).

\bibitem{Rapp03} R. Rapp, Nucl. Phys. {\bf A725}, 254 (2003);
S. Pratt and W. Bauer, Phys. Rev. {\bf C68}, 064905 (2003).

\bibitem{Becat04} F. Becattini, M. Ga\'{z}dzicki, A. Ker\"{a}nen, J. Manninen, R. Stock,
Phys. Rev. {\bf C69}, 024905 (2004); F. Becattini, J. Manninen, M. Ga\'{z}dzicki, Phys. Rev. {\bf C73},
044905 (2006) (hep-ph/0511092).

\bibitem{Ant06} N. G. Antoniou, F. K. Diakonos, A. S. Kapoyannis, K. S. Kousouris, Phys. Rev. Lett. {\bf 97}, 032002 (2006).

\bibitem{LoI} N. Antoniou {\it et al.}, {\it Study of Hadron Production
in Collisions of Protons and Nuclei at the CERN SPS}, CERN-SPSC-2006-034, CERN-SPSC-P-330 (2006).

\bibitem{RHICW} Workshop: {\it Can we discover the QCD critical point at RHIC?},
Brookhaven National Laboratory, March 9-10, 2006, Proceedings of RIKEN-BNL Research Center Workshop, BNL-75692-2006; PHENIX Collaboration,
S. S. Adler {\it et al.}, Phys. Rev. {\bf C76}, 034903 (2007).

\end{thebibliography}
\end{document}